\documentclass[pra,twocolumn,preprintnumbers,amsmath,amssymb]{revtex4}
\usepackage[T1]{fontenc}
\usepackage[latin2]{inputenc}

\usepackage{amsmath}
\usepackage{latexsym}
\usepackage{bbm}

\usepackage{graphicx}

\usepackage{color}

\def\dt{{\rm d}\,}

\def\duzomniejsze{<\kern-.7mm<}
\def\duzowieksze{>\kern-.7mm>}

\def\textbf#1{{\bf #1}}
\def\beq{\begin{equation}}
\def\eeq{\end{equation}}
\def\be{\begin{equation}}
\def\ee{\end{equation}}
\def\ben{\begin{eqnarray}}
\def\een{\end{eqnarray}}
\def\beqa{\begin{eqnarray}}
\def\eeqa{\end{eqnarray}}
\def\eea{\end{array}}
\def\bea{\begin{array}}
\newcommand{\bei}{\begin{itemize}}
\newcommand{\eei}{\end{itemize}}
\newcommand{\bee}{\begin{enumerate}}
\newcommand{\eee}{\end{enumerate}}

\def\1{\openone}

\def\tr{{\rm Tr}}

\def\>{\rangle}
\def\<{\langle}

\def\dt#1{{{\kern -.0mm\rm d}}#1\,}
\def\squareforqed{\hbox{\rlap{$\sqcap$}$\sqcup$}}
\def\qed{\ifmmode\squareforqed\else{\unskip\nobreak\hfil
\penalty50\hskip1em\null\nobreak\hfil\squareforqed
\parfillskip=0pt\finalhyphendemerits=0\endgraf}\fi}

\def\ep{\epsilon}

\newtheorem{lemma}{Lemma}

\newtheorem{theorem}[lemma]{Theorem}
\newtheorem{proposition}[lemma]{Proposition}
\newtheorem{definition}[lemma]{Definition}

\newtheorem{fact}[lemma]{Fact}

\def\bep{\begin{proposition}}
\def\eep{\end{proposition}}
\def\bel{\begin{lemma}}
\def\eel{\end{lemma}}

\def\bet{\begin{theorem}}
\def\eet{\end{theorem}}
\def\bed{\begin{definition}}
\def\eed{\end{definition}}
\def\bef{\begin{fact}}
\def\eef{\end{fact}}

\def\tep{{\tilde\epsilon}}
\def\t{\times}

\begin{document}

\title{Contextuality offers device-independent security}

\author{Karol Horodecki$^{1,2}$,
Micha\l{} Horodecki$^{3,2}$, Pawe\l{} Horodecki$^{4,2}$, \\Ryszard Horodecki$^{3,2}$, Marcin Paw\l{}owski$^{3,2}$
and Mohamed Bourennane$^{5}$}

\affiliation{$^{1}$Institute of Informatics, University of Gda\'nsk, 80-952 Gda\'nsk, Poland}

\affiliation{$^{2}$ National Quantum Information Centre of Gda\'nsk, 81-824 Sopot,
Poland}

\affiliation{$^{3}$ Institute of Theoretical Physics and Astrophysics, University of Gda\'nsk, 80-952 Gda\'nsk, Poland}

\affiliation{$^{4}$ Faculty of Applied Physics and Mathematics,Gda{\'n}sk University of Technology, 80-952 Gda{\'n}sk, Poland}

\affiliation{ $^{5}$ Department of Physics, Stockholm University, SE-10691 Stockholm, Sweden}

\begin{abstract}
The discovery of quantum key distribution by Bennett and Brassard (BB84) 
bases on the fundamental quantum feature: {\it incompatibility} of measurements 
of quantum non-commuting observables. 
In 1991 Ekert showed that cryptographic key can be generated at a distance 
with help of entangled (correlated) quantum particles.
Recently Barrett, Hardy and Kent showed that the \textit{non-locality} 
used by Ekert is itself a good resource of cryptographic key even beyond 
quantum mechanics. Their result paved the way to new generation
of quantum cryptographic protocols - secure even if the devices are built by the
very eavesdropper. However, there is a question, which is fundamental from both practical and philosophical point of view: does Nature offer security on operational
level based on the original concept behind quantum cryptography -
that information gain about one observable must cause disturbance to another, 
incompatible one?

Here we resolve this problem by using  another striking feature of quantum 
world - {\it contextuality}. It is a
strong version of incompatibility manifested in the famous Kochen-Specker
paradox. The crucial concept is the use of a new class of families of 
bipartite probability distributions which
locally exhibit the Kochen-Specker paradox conditions and, in addition, exhibit
perfect correlations. We show that if two persons share systems described by
such a family then they can extract secure key, even if they do not 
trust the devices which produce the statistics.  This is the first operational protocol that directly implements the fundamental
feature of Nature: the information gain vs. disturbance trade-off. 

For sake of proof we exhibit a new  Bell's inequality which is interesting in itself.
The security is proved not by exploiting strong violation of the inequality 
by quantum mechanics (as one usually proceeds), but rather by arguing, that quantum mechanics cannot violate it too much.

\end{abstract}

\maketitle

While quantum mechanics has been well established basis for modern technology
for years, recently we faced completely new possibilities which quantum mechanics
offers for processing of information. One of the landmarks is the
quantum cryptography \cite{Wiesner,BB84}, which allows to obtain secure cryptographic
key. The first quantum protocol for key distribution was given by Bennett and
Brassard in 1984 \cite{BB84}, and it bases on fundamental  quantum mechanical trade-off between
information gain and disturbance. Couple years later Ekert proposed a new idea for
quantum cryptography based on peculiar quantum correlations which may be shared by
distant particles called {\it entanglement} \cite{Ekert91}.

The seminal paper by Ekert carried actually two quite different concepts: (i) that
quantum {\it nonlocality} can be responsible for secure key and (ii) that
the information gain vs. disturbance trade-off of BB84 can be expressed in
terms of entanglement. The latter idea, clarified by Bennett, Brassard
and Mermin (BBM) \cite{BBM92} quite quickly turned out to be crucial for the field  of quantum
cryptography. In contrast, the first concept has become fashionable only much later:
after pioneering paper by Barrett, Hardy and Kent \cite{BHK_Bell_key} it became a boost for a new generation of cryptographic  protocols \cite{masanes-2006,masanes-2009-102,Nonsig_theories,hanggi-2009} - those exploiting solely the resource of non-locality, without referring to quantum mechanics at all.
The protocols of the new generation are now of central interest, as they may become crucial for the modern technology. Indeed, they pave the way to device-independent cryptography (i.e. when the devices for producing  secret key  may be produced by the very eavesdropper
\cite{masanes-2006,acin-2006-8,acin-2007-98,masanes-2009-102,Nonsig_theories,hanggi-2009}). There are essentially two approaches to device-independent 
security: either one assumes solely  no-signaling, or one assumes validity of quantum mechanics.  In either case, devices are not trusted, and 
security is verified solely through the statistics of the 
measurement outcomes. 

Remarkably, BB84 as well as its entanglement based version BBM have 
never received a device-independent extension. A fundamental question arises:
can the phenomenon of information gain vs. disturbance trade-off
be used to run device-independent cryptography?
At the first sight it could seem that the situation is hopeless:
as we will see below, there is a serious obstacle to make the BB84 protocol 
device-independent. One could therefore think, that
the main concept behind BB84 - the above mentioned trade-off -
cannot be put to work without complete specification of 
the quantum device.

\begin{figure}
\includegraphics [width=8cm]{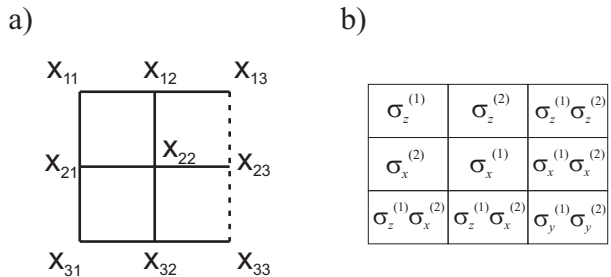}
\caption{\label{fig:Peres-Mermin} Peres-Mermin version of Kochen-Specker paradox.
We have 9 observables $x_i$ arranged into $3\times 3$ array.
If one chooses the observables as in (b) - where
we have two two-level systems and $\sigma^{(i)}$'s
are Pauli matrices on $i$-th system -
quantum mechanics allows for joint measurement only of observables
in a chosen row or a chosen column. One can ask whether some better theory could
reproduce quantum mechanical predictions, allowing however to predict
outcomes of all nine observables at the same time. This was the subject
of the famous Einstein-Bohr controversy. The Kochen-Specker paradox says that it is impossible. Namely, quantum mechanics predicts that along solid lines, the outcomes, if multiplied give with certainty 1, while on the dashed line they give -1.
Thus, supposing that these nine observables have some preexisting values,
which are merely revealed by measurement, we would obtain different value
of the product of all nine of them, if multiply them in different order,
which is a contradiction. So if one insists on ascribing some definite
values to observables, the value of at least one of them  would need to depend on whether the given observable  is measured within row or within column, i.e. on the context.  Thus only {\it contextual} values can be ascribed.  
Recently, Kochen-Specker paradox was expressed
in terms of inequalities \cite{Cabello-independent} which paved
the way to experimental verification of contextuality. The latter, 
however, is still not fully operational and needs some additional assumptions.
}
\end{figure}

In this paper  we argue that the obstacle can be overcome, by use of BBM protocol,
a version which was initially thought to be formally equivalent to BB84.
The main resource that allows to make
the trade-off operational, and use it for cryptography is {\it contextuality}
\footnote{The Kochen-Specker paradox has been considered in the context of
quantum secure key, see e.g. \cite{BechmanPeres-2000,nagata-2005-72,Svozil}
yet this was not done within the device-independent regime}. It is manifested by the famous Kochen Specker paradox \cite{Kochen-Specker},
which  received recently much attention being developed both theoretically
\cite{Spekkens-preparations,Spekkens-contextuality,Cabello-independent,Badziag-universal,guhne-2009} and tested experimentally \cite{michler-2000-84,kirchmair-2009-460,bartosik-2009-103,Amselem-independent,Liu-noncontextual,moussa-2009}. However, it should be noted, that all those experiments 
require some additional assumptions, which cannot be operationally verified 
(see \cite{guhne-2009}).
We shall refer to most popular version of KS paradox - the Peres-Mermin one \cite{Mermin1990-KS,Peres1990-KS} (see Fig. \ref{fig:Peres-Mermin}). However the present approach seems to be quite general and suitable for other variants of the paradox.

Since our result takes as a starting point the BBM protocol, it carries out the whole philosophy behind BB84 with one important difference. Namely, in
the proposed  protocol Alice makes measurement on a system which will never be in hands of Eve afterwards. This feature is crucial, if we want to have operational scheme, i.e. the device-independent one.
Indeed,  suppose that Alice measures a system which then goes into Eve's hands.
Then the system may simply carry the information
which measurement of Alice was performed \footnote{We enjoyed
discussion with Debbie Leung and Andreas Winter on this subject.}.

Our protocol is instead based on pairs of systems,
one kept by Alice and one sent to Bob. This prevents from imprinting
the information "which measurement" into the system
to which Eve has access. Therefore, the proposed protocol is as close as possible
to BB84, but not more.

Let us emphasize, that although our goal is to obtain security 
without direct referring to non-locality itself, our protocol must somehow involve
non-locality. Indeed if, conversely, there existed some hidden variables,
Eve can possess them, hence knowing everything about Alice and Bob systems.
And in fact, there is a deep connection between Kochen-Specker paradox and 
Bell inequalities: contextuality being a heart of the KS-paradox 
translates into non-locality manifested by violation of Bell inequalities 
(see e.g. \cite{Stairs-1983,HeywoodR1983,Cabello2001}).
Interestingly, in our proof, instead of using directly the fact, 
that there is strong non-locality, we shall rather argue the opposite - 
that the system offers security, because quantum mechanics 
cannot allow for certain too-strong non-locality. 

As a by-product we obtain a new (to our knowledge) Bell's inequality,
having peculiar properties. Namely, a recently introduced new 
principle {\it information causality} \cite{pawlowski-2009-461} allows to violate it up to maximal algebraic bound, while quantum mechanics does not. The inequality may thus play important role in studying the fundamental question to what extent information causality reproduces quantum mechanical limitations of Nature. 

{\bf The essential features of BB84 .-} Let us briefly recall the BBM version of BB84
(in Lo-Chau-Ardehali style \cite{LoChauArdehali2000}).
Alice and Bob share many pairs in maximally entangled state.
They select a random sample for testing purposes.
On this sample, Alice and Bob measure on their particles at random one of two non-commuting observables
$\sigma_x$ and $\sigma_z$. In case the particles are photons, the observables
may be taken horizontal vs. vertical polarization and $45^\circ$ vs. $135^\circ$
polarization respectively.
If the choices of Alice and Bob agree, the outputs should be correlated
if the state is indeed maximally entangled.
In presence of Eavesdropper or noise, the state may not
be exactly maximally entangled anymore, and therefore Alice and Bob
observe some error rate (correlations are not perfect).
If the error rate is too large, they abort the protocol.
If not, they measure $\sigma_z$  on the remaining pairs.
The outcomes of the latter measurement are called
{\it raw key}. If the error were zero, then
they would constitute perfect key, while if the
error is nonzero, but not too large, Alice and Bob
can apply procedures called error correction and privacy amplification
to obtain (asymptotically) perfect key from the raw key.

Thus the main idea behind is that Alice and Bob
check if they  have correlations, and any Eavesdropper
must destroy the correlations, in order to gain knowledge.
Let us emphasize: we do not say here about special "non-local" correlations
but just standard correlations meaning that outcome of Alice's measurement 
is the same as outcome of Bob's measurement. Eve must introduce disturbance, because the correlations are observed  for outcomes of {\it noncommuting} observables - the ones that cannot be simultaneously measured.
This suggests that in our operational analogue
we should use Kochen-Specker paradox, which is a 
manifestation of impossibility to measure some observables jointly.

We shall consider a notion of a "box" \cite{Popescu-Rohrlich} - a family of probability distributions.
The box has finite number of inputs  and for given input,
it returns output, whose statistics is described
by respective probability distribution. The inputs are simply observables.
Our box will respect quantum theory, i.e. it will be physically realizable.
On the other hand, since the number of probability distributions
is finite, it can be tested statistically, without any knowledge
of how the device was built.

For the purpose of our protocol, we shall propose a {\it Kochen-Specker bipartite box}, which will exhibit the following features: on one hand the local outcomes will
satisfy KS constraints, which implies that they have to
come from observables that cannot be  measured jointly, and on the other hand
if the same observable is measured by Alice and Bob, the results
are perfectly correlated (see  \cite{conway-2007} in this context). We shall then prove, that such a box provides about half bit of secrecy. Then we will consider a noisy version  of the box.

{\bf Peres-Mermin version of the Kochen Specker paradox .-} We shall use
the Peres-Mermin version of  KS paradox \cite{Mermin1990-KS,Peres1990-KS}. 
The quantum observables and the KS conditions are depicted on Fig. \ref{fig:Peres-Mermin}.
The Peres-Mermin box (PM box) is the set of 6 joint probability distributions.
Namely, the box has nine observables (inputs) $x_{ij},\, i,j\in\{1,...,3\}$ in $3\times 3$ array, and
in any given row (column) the binary observables can be measured at the same time.
The box is therefore a family of six probability distributions,
(3 rows and 3 columns) and each distribution is a joint distribution
of three observables. We demand that the outputs satisfy the following condition:
\bei
\item{}[KS condition] The 6 joint distributions satisfy constraints, coming from Kochen-Specker type paradox: measuring the rows and two leftmost columns, one always get even number of +1's, while the last column has always odd number of +1's.
\eei

In the following we shall consider a distributed version of the above box.

{\bf Ideal distributed PM box and intrinsic randomness.-} We define a distributed Peres-Mermin box, shared by Alice and Bob as follows.  
Both Alice and Bob have Peres-Mermin array of observables,
which locally satisfy the above mentioned conditions (KS and compatibility ones). 
Alice measures columns of the array, while Bob measures rows. 
This, in particular assures, that e.g. at Alice's site 
each observable from the array is measured 
in a fixed context, unlike in original KS paradox, 
where there is only one laboratory, and observables 
have to be measured in two different contexts. The same holds for Bob's site.
In addition we assume
{\it AB-correlations}, i.e. that there are perfect correlations between the outcomes of the same observables on Alice and Bob side.
Also, we consider a non-signaling, meaning, that Alice's local  distributions do not depend on the choice of measurement by Bob. The no-signaling condition allows to say meaningfully about Alice's and Bob's local distributions (i.e. allow for existence of {\it subsystems}). Let us emphasize, that the distributed version of Peres-Mermin box 
exhibit necessarily non-locality (as is actually true  for distributed 
version of any KS paradox see e.g. \cite{Stairs-1983}). Indeed, in distributed scenario non-contextuality translates into  non-locality, which in turn is 
a necessary condition for security. 

\begin{figure}
\includegraphics [width=7cm]{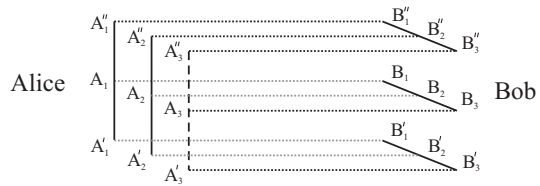}
\caption{\label{fig:DPM-box} The distributed Peres-Mermin box. Solid or dotted line means that there is an even number of  $-1$'s while dashed line  - odd number of $-1$'s.}
\end{figure}

Formally, distributed PM box is a family of 9 conditional distributions $P({\bf a},{\bf b}|A,B)$
Here $A=1,2,3$ runs over columns of PM array, $B=1,2,3$ runs over  rows of the array. and ${\bf a}=(a_1 a_2 a_3)$ and  ${\bf b}=(b_1 b_2 b_3)$ denote triples of outcomes.  
As said, the family has to satisfy the following conditions:
\bei
\item{}[KS condition] For $A=1,2$ and $B=1,2,3$ 
the product of outcomes is 1, i.e. ${\bf a}\in\{+++,--+,-+-,--+\}$, and the same for ${\bf b}$. For $A=3$ the outcomes with nonzero probability 
multiply up to $-1$, so that  ${\bf a}\in\{---,-++,+-+,++-\}$ in this case. 
\item{}[AB correlations] For $A=i$ and $B=j$ we have  $a_i=b_j$.
\item{}[no-signaling] The marginal probability $P({\bf a}|A,B)$ does not depend on $B$ and similarly for $P({\bf b}|A,B)$ does not depend on $A$.
\eei
Further we shall use a different notation, depicted on Fig. \ref{fig:DPM-box}.

{\bf Intrinsic randomness from ideal distributed PM box.-}
We shall now assume that a distributed PM box comes from quantum mechanics,
but we do not know how it is implemented (i.e. what observables 
are measured, and in which quantum state). 
We are thus in a paradigm of {\it device-independent} security 
\cite{acin-2007-98} which assumes validity of quantum mechanics. Under such assumption
we shall show that the outcomes of a fixed row/column (for definiteness
take first row of Bob's system) possess 
about 0.44 bits of intrinsic randomness (hence they offer security).
Consider first a simpler problem, namely let us prove, that
on Bob's side, the outcomes of first row and first column cannot 
be all deterministic, i.e. their marginal probabilities 
cannot be all $0$ or $1$). Suppose, conversely, 
that they are all deterministic. 
Without loss of generality, we can assume that all the involved observables 
($B_1',B_1,B_1'',B_2'',B_3''$) have value $+1$. 
Due to perfect correlations the corresponding $A$'s are also set to $+1$. 
Exploiting now AB-correlations and KS-condition, one obtains
super-strong correlations for the following four outcomes:
$a\equiv A_2, a'\equiv A_3',b\equiv B_2', b'\equiv B_3$.
Namely, pairs $(ab),(ab'),(a'b)$ are perfectly correlated,
while $(a'b')$ is perfectly anti-correlated. Such correlations 
would violate the CHSH inequality \cite{CHSH}
\be
|\<ab\>+\<ab'\>+\<a'b\>-\<a'b'\>|\leq 2
\ee
up to 4, while it is well known 
that quantum mechanics, due to Tsirelson bound \cite{Tsirelson-bound}
allows only for $2\sqrt 2$.

\begin{figure}
\includegraphics [width=7cm]{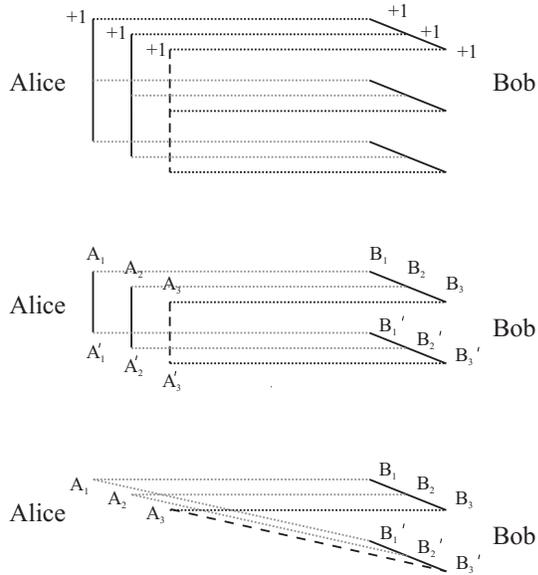}
\caption{\label{fig:reduction}Deterministic values for first Bob's row lead to 
a system of perfect correlations and anticorrelations violating maximally 
a Bell inequality. Solid or dotted line means that there is an even number of 
$-1$'s while dashed line  - odd number of $-1$'s.}
\end{figure}

Let us now show that also the first row itself cannot have 
deterministic values at Bob's side. Assuming now that 
the values in the row are all $+1$, and exploiting KS-condition and 
AB-correlations, we shall now obtain a system of correlations.
Note, that due to KS condition, we have $B_3=B_1B_2$,
and $B_3'=B_1'B_2'$. (We cannot say that same about $A$'s 
because they are not measured in rows). 
Now, values $+1$ in first row imply, that in addition to perfect 
correlations for pairs $(A_i,B_i)$ with $i=1,2,3$, 
we have also perfect correlations for pairs $(A_1,B_1')$ and $(A_2,B_2')$
and perfect anti-correlations for $(A_3,B_3')$. The whole reasoning 
is depicted on Fig. \ref{fig:reduction}.
This means that they violate the following  Bell inequality 
\ben
&&\gamma(A:B) \equiv \<A_1 B_1\> + \<A_2 B_2\>+\<A_3 B_3\>\nonumber \\
&&+\<A_1 B_1'\> +\<A_2 B_2'\>-\<A_3 B_3'\>\leq 4
\label{eq:Bell-gamma}
\een
(where $B_3\equiv B_1B_2$, $B_3'\equiv B_1'B_2'$)
up to 6, i.e. the correlations again reach the absolute, algebraic bound.  
Now, one can show that this is impossible from quantum mechanics. 
Namely, it was shown in \cite{GisinMS2006-pseudo} we know, that in quantum mechanical 
correlations cannot allow to win a so-called "pseudo-telepathy" game
\footnote{Pseudo-telepathy game \cite{Brassard-pseudo} is a generalization 
of GHZ-type paradoxes.}
which cannot be won within a classical theory, if the game  involves 
no more than two observables on one of the sites.  In our case, since $B_i$'s  
can be jointly measured, they can be regarded as a single observable 
with four outcomes (same for $B_i'$).  Thus we have two observables on Bob's site, and one can show that so extremal violation of the above Bell inequality actually means winning a certain pseudo-telepathy game. We shall 
not describe it in a full detail here, because we intend to provide 
a quantitative upper bound for quantum mechanical violation 
of the inequality. However, already at this point, we can notice
that while usually, to prove device-independent security 
one was looking for a large violation, in our case, we have to do the converse 
job, and rather show that some Bell's {\it cannot be} violated too strongly by 
quantum mechanics.

So far we have argued, that the values in first row  cannot be deterministic. However, in order to use it  for cryptography, we need some quantitative statements. 

Let us thus first relate the degree of violation of the inequality 
and  the constraints on probability distribution of the first row. 
To this end we shall use an equivalent inequality 
expressed in terms of probabilities rather than correlations:
\ben
&&\beta(A:B)\equiv p(A_1=B_1)+p(A_2=B_2)+p(A_3=B_3)\nonumber \\
&&+p(A_1=B_1')+p(A_2=B_2')+p(A_3\not=B_3')\leq 5.
\label{eq:beta}
\een
One finds that $\beta(A:B)=(\gamma(A:B)+6)/2$. 
Let $p_i=p(i|+)$ be probability that $b_{1i}=+1$,
i.e. it is marginal probability of obtaining $+1$ for a given outcome 
from the first row on Bob's site. Then  by using  elementary inequality $p(C\cap D)\geq p(C)+p(D) -1$ for any events $C$ and $D$, and exploiting AB-correlations and KS-condition we obtain that 
\be
\beta(A:B)=p_1+p_2+p_3 +3.
\ee
Suppose now we have the following bound for quantum mechanical 
violation of inequality \eqref{eq:beta}: 
$\beta(A:B)\leq \beta_0<6$. We obtain 
\be
p_1+p_2+p_3 \leq \beta_0-3
\ee
i.e. not all of  $p_i$ can be 1. In a similar way, employing 
three other possibilities of assigning deterministic values to 
the Bob's first row ($+--$,$-+-$ and $--+$)
one obtains
\ben
&& p_1+ (1-p_2)+(1-p_3) \leq \beta_0-3\nonumber\\
&& (1-p_1)+ p_2+(1-p_3) \leq \beta_0-3\nonumber\\
&& (1-p_1)+ (1-p_2)+p_3 \leq \beta_0-3
\label{eq:cons-p}
\een
Let us now consider joint probability distributions 
for the outcomes of the first row on Bob's site $q_0=q(+++), q_1=q(+--),
q_2=q(-+-),q_3=q(--+)$, with $q_0+q_1+q_2+q_3=1$.  One gets the following 
relation 
\ben
&&q_0= \frac12(-1+ p_1 +p_2+p_3)\nonumber \\
&&q_1= \frac12(1+ p_1 -p_2-p_3)\nonumber \\
&&q_2= \frac12(1+ p_2 -p_2-p_3)\nonumber\\
&&q_3= \frac12(1+ p_3 -p_1-p_2)
\een
Then the constraints \eqref{eq:cons-p} imply 
\be
q_i\leq\frac12(\beta_0-4) =\frac14(\gamma_0-2)
\label{eq:cons-q-gamma}
\ee
for $i=0,1,2,3$. We have obtained numerically $\gamma_0=5.6364$,  which gives 
\be
q_i\leq x.
\label{eq:cons-q}
\ee
with $x\lesssim 0.9091$. 
Let us note here is easy to see that $x \geq 1/2$,
as this value would correspond to $\beta_0=5$,
clearly achievable even within classical theory. Finally note, 
that due to AB-correlations, we have the same bounds for probabilities in 
Alice's first row.

{\bf Secure key from ideal box.-}
Suppose that Alice and Bob share a bipartite box $R_{AB}$ (which they can verify
by testing samples). This box may be decomposable into some other boxes,
\be
R_{AB}=\sum_e q_e R^e_{AB}
\label{eq:ensemble}
\ee
(where we deliberately label boxes by $e$,
as it will be Eve, who knows which box $R^e_{AB}$ is actually realized).
The decomposition \eqref{eq:ensemble} arises as follows:
Eve creates a joint box $R_{ABE}$, and hands the part $R_{AB}$
to Alice and Bob. When they announce their choices of measurements,
Eve measures her part, and we do not specify her inputs and outputs 
(the latter denoted by $e$), so that 
her power is to split the Alice and Bob box $R_{AB}$ into arbitrary
ensemble $\{q_e, R^e_{AB}\}$ which satisfies  $ \sum_e q_e R^e_{AB}=R_{AB}$. 
This power is analogous to the situation in quantum mechanics, where we hand to Eve 
the purification of the Alice and Bob state, which means that Eve controls the whole 
Universe, except of Alice's and Bob's labs. That this is the only thing which Eve can do, follows from impossibility of signaling from Eve to Alice and Bob (see 
\cite{acin-2006-8}).

Now, let the box $R_{AB}$ be a bipartite Peres-Mermin  box.
Then the  boxes $R_{AB}^e$ must be bipartite Peres-Mermin ones, too.
This is because the conditions determining bipartite PM boxes
(i.e. KS condition and AB-correlations) are formulated as ascribing
certain probabilities value 0, while no-signaling is a principle 
assumed to be always true (as  we shall eventually use quantum mechanics
which obviously obeys no-signaling). 
Thus Eve can only decompose the box $R$ again into distributed PM-boxes.

Suppose now Alice and Bob share $n$ boxes.
They select a sample to verify, that they share distributed PM box.
In verification procedure, Alice measures at random columns,
while Bob measures at random rows. From the rest of boxes 
they would like to draw key.  We have shown, that 
Eve cannot know the first row exactly,
while Alice and Bob are correlated. So Alice and Bob 
should both measure first row,
and thanks to AB-correlations, would obtain key.
However if Alice would measure row, she has to 
use a different setup of the device, 
than she used when measuring columns. 
Since our protocol is to be device-independent,
we have to assume, that the device can be malicious. 
In particular, when Alice wants to measure rows, she may in fact measure 
some completely different observables
than when she measured columns. And Eve can know the outcomes 
of those new observables perfectly. 

However, since we already know, that outcomes of Bob's first row 
are relatively secure, it is enough that Alice measures 
also the first row provided, that the outcomes 
are perfectly correlated with the Bob's row, which 
now we assume. 



This suggests the following protocol.
Alice and Bob share $n$ pairs of particles. They select two samples.
On the first sample, they measure at random columns and rows, respectively. 
This allows to verify that they indeed share a distributed PM box. 
On the second sample  Alice and Bob measure just the first row,
and verify whether their outcomes are correlated.  On the rest of pairs 
Alice and Bob measure also the first row, 
and the outcomes constitute a so called {\it raw} key. 
Then they apply standard procedures of error correction and privacy amplification,
to obtain shorter, but secure key. 
More precisely: in the case of ideal PM box, error correction 
is not needed (as the outcomes of Alice and Bob are by perfectly correlated) and only privacy amplification will be performed. The error correction is 
be needed in the noisy case, which we will further describe.


To estimate the amount of obtained secure key, we shall now analyse the 
triple of random variables $(A,B,E)$.
The $A$ and $B$ are variables describing the first row of Alice and Bob respectively,
and $E$  is the variable of Eve describing the choice of ensemble (\ref{eq:ensemble})
i.e. $E=e$ with probability $q_e$. Now we can use  the well 
known Csiszar-K\"{o}rner formula 
\be
K\geq I(A:B)-I(A:E)
\label{eq:CK}
\ee 
which provides a 
lower bound for the rate $K$ of secure key \cite{CsisarKorner_key_agreement,Maurer_key_agreement} 
where $I(A:B)$ denotes Shannon mutual information of the joint probability 
distribution of Alice's and Bob's first row, 
and $I(A:E)$ denotes mutual information of 
joint probability distribution of Alice's row 
and output of Eve's measurement. The formula  rewrite the expression by means of conditional entropies: $H(A|E) - H(A|B)$. Since Alice and Bob are perfectly correlated, $H(A|B)=0$. However 
Eve will split the box into an ensemble, any member of ensemble 
is PM box has to satisfy the bound \eqref{eq:cons-q}. 
One easily find, that the distribution of smallest entropy,
which satisfies the bound is $(x,1-x,0,0)$. Thus $H(A|E)\geq 0.439$, and 
this rate of secure key can be obtained.  

The considered ideal distributed PM box corresponds to the situation,
where there is no disturbance (perfect 
correlations between Alice and Bob). We see that in this case,  Eve can gain some knowledge, however she cannot have full knowledge. We thus observe a version of information-gain vs. disturbance trade-off, where it is impossible 
to gain full knowledge about the system without disturbing 
its correlations with another system.  Below we will complete the picture
by considering presence of disturbance.

{\bf Noisy case .-}
Suppose that Alice and Bob while measuring correlations
obtain average bit error rate $q$ (i.e. $q$ is the ratio
of anti-correlated events to all events). First of all,
let us note that the noisy box can still satisfy
perfectly KS conditions. This is
because Alice and Bob can force it by measuring only two observables
out of three (that forms say a row), and fabricating the outcome of 
the third one, rather than measuring it. Therefore, the noise will 
influence correlations between Alice and Bob. We shall 
assume that on the test samples, for any observable 
from the Peres-Mermin array Alice and Bob  
obtained correlations with probability $1-\ep$, i.e. $\ep$ 
measures the error level. Clearly, the larger the error, 
the looser are the constraints for probabilities of the Bob's first row. 
One can find (see Appendix) that the constraints \eqref{eq:cons-q} become now
\be
q_i\leq x- 4.5 \ep 
\ee
with $x\simeq0.9091$.
From this, one can obtain the following bound on $H(E|B)$:
\be
H(B|E)\geq \sup_{\delta>0}(1-\frac{\ep}{\delta})h(x +4.5\delta).
\label{eq:hbe}
\ee
The lack of correlations influences also $H(B|A)$, 
which in turn can be bounded as follows 
\be
H(B|A)\leq h(\frac32 \ep)+\frac32 \ep \log 3.
\label{eq:hba}
\ee
Inserting it to the Csiszar-K\"orner formula \eqref{eq:CK},  and putting 
$\delta=1.8$ we obtain  that for $\ep \lesssim 0.68\%$ secure key can be obtained. This is smaller than typical thresholds obtained  from CHSH, which are of 
order of $2\%$. However our estimates have not all been tight, and there 
is still some room for optimization.

{\bf Quantum mechanical implementation and the security level.-}
So far we have considered an abstract box. To make use of our results, 
we need to know, that the box can be realized in labs, i.e. 
that the box can be simulated quantum mechanically. Indeed, it is the case:
the box is obtained by Alice and Bob measuring Peres-Mermin observables 
(see Fig. \ref{fig:Peres-Mermin}) on two pairs of qubits 
in maximally entangled state 
(which were recently used to derive non-locality from contextuality \cite{cabello-2009}, see also \cite{Cabello2001}).
Let us mention here, that if we had a tripartite PM box, 
with perfect tripartite correlations,
then one can show that one secure bit can be obtained
by the three parties. And moreover,
the security would come solely from no-signaling assumption. 
However, though there exists such a no-signaling box, unfortunately, cannot be implemented by quantum mechanical devices, i.e. it does not exist in Nature
(this is actually implied by our present result).


Our reasoning proves that the obtained key is secure under the 
so called individual attack: Eve couple to each box independently,
and measures before Alice and Bob perform 
error correction and privacy amplification procedures.  
 We believe that one can apply the ideas of \cite{masanes-2006,acin-2007-98,masanes-2009-102,hanggi-2009} to obtain stronger security.

{\bf Conclusions .-} We have provided
the first operational protocol that directly implements the fundamental
feature of Nature: the information gain vs. disturbance trade-off rather than
used so far non-locality. It implements the original
idea of BB84, that one who gains information, will at the same time
introduce disturbance, but does it on operational level, i.e.
the security is here verified by the very statistics of the
measurement,  allowing for the devices to be built by the very Eavesdropper.
The trade-off appears, because Alice and Bob measure 
incompatible quantities, which we imposed by applying Kochen-Specker paradox.


Let us compare our approach with the previous device-independent protocols
which directly employ non-locality.  Clearly our bipartite box must be non-local,
otherwise, as argued by Ekert, Eve could have full knowledge
of all results of measurements. However it is interesting to find a more direct
relation between our approach and non-locality approach. To this end,
one may employ non-locality obtained by Cabello \cite{cabello-2009}
precisely for the distributed PM paradox, 
and derive from it existence of the key to see the connection.
Paradoxically, we have proved security not by exploiting 
the fact that our system exhibits a strong non-locality, 
but rather by showing, that a part of our total system 
cannot exhibit too strong non-locality. As a by-product
we have obtained a new Bell's inequality, whose maximal violation is excluded 
only within so-called second hierarchy of necessary conditions 
for a given distribution be reproducible by quantum mechanics
according to classification of \cite{navascues-2008-10}. 

Remarkably, the exhibited bipartite box allows Eve's to gain some information
without causing any disturbance. Thus the trade-off is not so strict
as the one offered by full quantum formalism. However this information gain
is smaller than the amount of correlations shared by Alice and Bob (2 bits),
which allows for creating secure key.

The fact, that from KS paradox one can obtain security, appears to be
not occasional from yet another point of view. Namely, some sets of KS like observables (yielding paradox), lead also to some error correcting codes 
\cite{divincenzo-1997-55}, and the latter are again connected 
with security \cite{kretschmann-2008-78}.  

Finally, our work suggests some further developments. First of all,
in our paper we have obtained device-independent security, 
which assumes validity of quantum mechanics. There is an open question,
whether one can have a stronger version of device-independent security 
- the one based solely on no-signaling, and not assuming 
any knowledge about quantum mechanics. To this end, one should 
analyse other Kochen-Specker paradoxes and apply them to obtain security. 
In other words, we believe that analysis of restricted class 
of non-local scenarios - the ones being distributed versions of 
local Kochen-Specker paradoxes may lead to some interesting general results 
on device-independent security. 

\begin{acknowledgments}
M.P. thanks Esther H{\"a}nggi for helpful discussions. 
K.H., M.H. and R.H. thank Debbie Leung and Andreas Winter for helpful 
discussions at the early  stage of our work on the problem. 
We thank Adan Cabello, Stefano Pironio and Valerio Scarani for useful feedback and valuable references. We are grateful 
to Andrzej Grudka, Pawe\l{} Kurzy\'nski and Antoni W\'ojcik 
for critique of the earlier version of the paper,
which has lead to revealing a crucial error.
This work was supported by the European Commission through 
the Integrated Project FET/QIPC ``QESSENCE''. M.H., P.H. and R.H. 
are supported by Ministry of Science 
and Higher Education, grant N N202 231937. Part of this work was done in National Quantum Information Centre of Gda\'nsk.
\end{acknowledgments}

\bibliographystyle{apsrev}

\bibliography{rmp14-hugekey}

\begin{thebibliography}{46}
\expandafter\ifx\csname natexlab\endcsname\relax\def\natexlab#1{#1}\fi
\expandafter\ifx\csname bibnamefont\endcsname\relax
  \def\bibnamefont#1{#1}\fi
\expandafter\ifx\csname bibfnamefont\endcsname\relax
  \def\bibfnamefont#1{#1}\fi
\expandafter\ifx\csname citenamefont\endcsname\relax
  \def\citenamefont#1{#1}\fi
\expandafter\ifx\csname url\endcsname\relax
  \def\url#1{\texttt{#1}}\fi
\expandafter\ifx\csname urlprefix\endcsname\relax\def\urlprefix{URL }\fi
\providecommand{\bibinfo}[2]{#2}
\providecommand{\eprint}[2][]{\url{#2}}

\bibitem[{\citenamefont{Wiesner}(1983)}]{Wiesner}
\bibinfo{author}{\bibfnamefont{S.}~\bibnamefont{Wiesner}},
  \bibinfo{journal}{Sigact news} \textbf{\bibinfo{volume}{15:1}},
  \bibinfo{pages}{78} (\bibinfo{year}{1983}).

\bibitem[{\citenamefont{Bennett and Brassard}(1984)}]{BB84}
\bibinfo{author}{\bibfnamefont{C.~H.} \bibnamefont{Bennett}} \bibnamefont{and}
  \bibinfo{author}{\bibfnamefont{G.}~\bibnamefont{Brassard}}, in
  \emph{\bibinfo{booktitle}{Proceedings of the IEEE International Conference on
  Computers, Systems and Signal Processing}} (\bibinfo{publisher}{IEEE Computer
  Society Press, New York}, \bibinfo{address}{Bangalore, India, December 1984},
  \bibinfo{year}{1984}), pp. \bibinfo{pages}{175--179}.

\bibitem[{\citenamefont{Ekert}(1991)}]{Ekert91}
\bibinfo{author}{\bibfnamefont{A.~K.} \bibnamefont{Ekert}},
  \bibinfo{journal}{Phys. Rev. Lett.} \textbf{\bibinfo{volume}{67}},
  \bibinfo{pages}{661} (\bibinfo{year}{1991}).

\bibitem[{\citenamefont{Bennett et~al.}(1992)\citenamefont{Bennett, Brassard,
  and Mermin}}]{BBM92}
\bibinfo{author}{\bibfnamefont{C.~H.} \bibnamefont{Bennett}},
  \bibinfo{author}{\bibfnamefont{G.}~\bibnamefont{Brassard}}, \bibnamefont{and}
  \bibinfo{author}{\bibfnamefont{N.~D.} \bibnamefont{Mermin}},
  \bibinfo{journal}{Phys. Rev. Lett.} \textbf{\bibinfo{volume}{68}},
  \bibinfo{pages}{557} (\bibinfo{year}{1992}).

\bibitem[{\citenamefont{Barrett et~al.}(2005)\citenamefont{Barrett, Hardy, and
  Kent}}]{BHK_Bell_key}
\bibinfo{author}{\bibfnamefont{J.}~\bibnamefont{Barrett}},
  \bibinfo{author}{\bibfnamefont{L.}~\bibnamefont{Hardy}}, \bibnamefont{and}
  \bibinfo{author}{\bibfnamefont{A.}~\bibnamefont{Kent}},
  \bibinfo{journal}{Phys. Rev. Lett.} \textbf{\bibinfo{volume}{95}},
  \bibinfo{pages}{010503} (\bibinfo{year}{2005}).

\bibitem[{\citenamefont{Masanes
  et~al.}(2006{\natexlab{a}})\citenamefont{Masanes, Renner, Christandl, Winter,
  and Barrett}}]{masanes-2006}
\bibinfo{author}{\bibfnamefont{L.}~\bibnamefont{Masanes}},
  \bibinfo{author}{\bibfnamefont{R.}~\bibnamefont{Renner}},
  \bibinfo{author}{\bibfnamefont{M.}~\bibnamefont{Christandl}},
  \bibinfo{author}{\bibfnamefont{A.}~\bibnamefont{Winter}}, \bibnamefont{and}
  \bibinfo{author}{\bibfnamefont{J.}~\bibnamefont{Barrett}},
  \emph{\bibinfo{title}{Unconditional security of key distribution from
  causality constraints}} (\bibinfo{year}{2006}{\natexlab{a}}),
  \eprint{arXiv.org:quant-ph/0606049}.

\bibitem[{\citenamefont{Masanes}(2009)}]{masanes-2009-102}
\bibinfo{author}{\bibfnamefont{L.}~\bibnamefont{Masanes}},
  \bibinfo{journal}{Phys. Rev. Lett.} \textbf{\bibinfo{volume}{102}},
  \bibinfo{pages}{140501} (\bibinfo{year}{2009}), \eprint{arXiv.org:0807.2158}.

\bibitem[{\citenamefont{Masanes
  et~al.}(2006{\natexlab{b}})\citenamefont{Masanes, Acin, and
  Gisin}}]{Nonsig_theories}
\bibinfo{author}{\bibfnamefont{L.}~\bibnamefont{Masanes}},
  \bibinfo{author}{\bibfnamefont{A.}~\bibnamefont{Acin}}, \bibnamefont{and}
  \bibinfo{author}{\bibfnamefont{N.}~\bibnamefont{Gisin}},
  \bibinfo{journal}{Phys. Rev. A} \textbf{\bibinfo{volume}{73}},
  \bibinfo{pages}{012112} (\bibinfo{year}{2006}{\natexlab{b}}).

\bibitem[{\citenamefont{Hanggi et~al.}(2009)\citenamefont{Hanggi, Renner, and
  Wolf}}]{hanggi-2009}
\bibinfo{author}{\bibfnamefont{E.}~\bibnamefont{Hanggi}},
  \bibinfo{author}{\bibfnamefont{R.}~\bibnamefont{Renner}}, \bibnamefont{and}
  \bibinfo{author}{\bibfnamefont{S.}~\bibnamefont{Wolf}},
  \emph{\bibinfo{title}{Efficient quantum key distribution based solely on
  bell's theorem}} (\bibinfo{year}{2009}), \eprint{arXiv.org:0911.4171}.

\bibitem[{\citenamefont{Acin et~al.}(2006)\citenamefont{Acin, Massar, and
  Pironio}}]{acin-2006-8}
\bibinfo{author}{\bibfnamefont{A.}~\bibnamefont{Acin}},
  \bibinfo{author}{\bibfnamefont{S.}~\bibnamefont{Massar}}, \bibnamefont{and}
  \bibinfo{author}{\bibfnamefont{S.}~\bibnamefont{Pironio}},
  \bibinfo{journal}{New J. Phys.} \textbf{\bibinfo{volume}{8}},
  \bibinfo{pages}{126} (\bibinfo{year}{2006}), \eprint{arXiv:quant-ph/0605246}.

\bibitem[{\citenamefont{Acin et~al.}(2007)\citenamefont{Acin, Brunner, Gisin,
  Massar, Pironio, and Scarani}}]{acin-2007-98}
\bibinfo{author}{\bibfnamefont{A.}~\bibnamefont{Acin}},
  \bibinfo{author}{\bibfnamefont{N.}~\bibnamefont{Brunner}},
  \bibinfo{author}{\bibfnamefont{N.}~\bibnamefont{Gisin}},
  \bibinfo{author}{\bibfnamefont{S.}~\bibnamefont{Massar}},
  \bibinfo{author}{\bibfnamefont{S.}~\bibnamefont{Pironio}}, \bibnamefont{and}
  \bibinfo{author}{\bibfnamefont{V.}~\bibnamefont{Scarani}},
  \bibinfo{journal}{Phys. Rev. Lett.} \textbf{\bibinfo{volume}{98}},
  \bibinfo{pages}{230501} (\bibinfo{year}{2007}),
  \eprint{arXiv.org:quant-ph/0702152}.

\bibitem[{\citenamefont{Kochen and Specker}(1967)}]{Kochen-Specker}
\bibinfo{author}{\bibfnamefont{S.}~\bibnamefont{Kochen}} \bibnamefont{and}
  \bibinfo{author}{\bibfnamefont{E.~P.} \bibnamefont{Specker}},
  \bibinfo{journal}{J. Math. Mech.} \textbf{\bibinfo{volume}{17}},
  \bibinfo{pages}{59} (\bibinfo{year}{1967}).

\bibitem[{\citenamefont{Spekkens}(2005)}]{Spekkens-preparations}
\bibinfo{author}{\bibfnamefont{R.~W.} \bibnamefont{Spekkens}},
  \bibinfo{journal}{Phys. Rev. A} \textbf{\bibinfo{volume}{71}},
  \bibinfo{pages}{052108} (\bibinfo{year}{2005}), \eprint{quant-ph/0406166}.

\bibitem[{\citenamefont{Spekkens et~al.}(2009)\citenamefont{Spekkens, Buzacott,
  Keehn, Toner, and Pryde}}]{Spekkens-contextuality}
\bibinfo{author}{\bibfnamefont{R.~W.} \bibnamefont{Spekkens}},
  \bibinfo{author}{\bibfnamefont{D.~H.} \bibnamefont{Buzacott}},
  \bibinfo{author}{\bibfnamefont{A.~J.} \bibnamefont{Keehn}},
  \bibinfo{author}{\bibfnamefont{B.}~\bibnamefont{Toner}}, \bibnamefont{and}
  \bibinfo{author}{\bibfnamefont{G.~J.} \bibnamefont{Pryde}},
  \bibinfo{journal}{Phys. Rev. Lett.} \textbf{\bibinfo{volume}{102}},
  \bibinfo{pages}{010401} (\bibinfo{year}{2009}), \eprint{quant-ph/0805.1463}.

\bibitem[{\citenamefont{Cabello}(2008)}]{Cabello-independent}
\bibinfo{author}{\bibfnamefont{A.}~\bibnamefont{Cabello}},
  \bibinfo{journal}{Phys. Rev. Lett.} \textbf{\bibinfo{volume}{101}},
  \bibinfo{pages}{210401} (\bibinfo{year}{2008}), \eprint{quant-ph/0808.2456}.

\bibitem[{\citenamefont{Badzi{\c{a}}g et~al.}(2009)\citenamefont{Badzi{\c{a}}g,
  Bengtsson, Cabello, and Pitowsky}}]{Badziag-universal}
\bibinfo{author}{\bibfnamefont{P.}~\bibnamefont{Badzi{\c{a}}g}},
  \bibinfo{author}{\bibfnamefont{I.}~\bibnamefont{Bengtsson}},
  \bibinfo{author}{\bibfnamefont{A.}~\bibnamefont{Cabello}}, \bibnamefont{and}
  \bibinfo{author}{\bibfnamefont{I.}~\bibnamefont{Pitowsky}},
  \bibinfo{journal}{Phys. Rev. Lett.} \textbf{\bibinfo{volume}{103}},
  \bibinfo{pages}{050401} (\bibinfo{year}{2009}), \eprint{quant-ph/0809.0430}.

\bibitem[{\citenamefont{Guhne et~al.}(2009)\citenamefont{Guhne, Kleinmann,
  Cabello, Larsson, Kirchmair, Zahringer, Gerritsma, and Roos}}]{guhne-2009}
\bibinfo{author}{\bibfnamefont{O.}~\bibnamefont{Guhne}},
  \bibinfo{author}{\bibfnamefont{M.}~\bibnamefont{Kleinmann}},
  \bibinfo{author}{\bibfnamefont{A.}~\bibnamefont{Cabello}},
  \bibinfo{author}{\bibfnamefont{J.~A.} \bibnamefont{Larsson}},
  \bibinfo{author}{\bibfnamefont{G.}~\bibnamefont{Kirchmair}},
  \bibinfo{author}{\bibfnamefont{F.}~\bibnamefont{Zahringer}},
  \bibinfo{author}{\bibfnamefont{R.}~\bibnamefont{Gerritsma}},
  \bibnamefont{and} \bibinfo{author}{\bibfnamefont{C.~F.} \bibnamefont{Roos}},
  \emph{\bibinfo{title}{Compatibility and noncontextuality for sequential
  measurements}} (\bibinfo{year}{2009}), \eprint{arxiv:0912.484}.

\bibitem[{\citenamefont{Michler et~al.}(2000)\citenamefont{Michler, Weinfurter,
  and Zukowski}}]{michler-2000-84}
\bibinfo{author}{\bibfnamefont{M.}~\bibnamefont{Michler}},
  \bibinfo{author}{\bibfnamefont{H.}~\bibnamefont{Weinfurter}},
  \bibnamefont{and} \bibinfo{author}{\bibfnamefont{M.}~\bibnamefont{Zukowski}},
  \bibinfo{journal}{Phys. Rev. Lett.} \textbf{\bibinfo{volume}{84}},
  \bibinfo{pages}{5457} (\bibinfo{year}{2000}),
  \eprint{arXiv:quant-ph/0009061}.

\bibitem[{\citenamefont{Kirchmair et~al.}(2009)\citenamefont{Kirchmair,
  Zahringer, Gerritsma, Kleinmann, Guhne, Cabello, Blatt, and
  Roos}}]{kirchmair-2009-460}
\bibinfo{author}{\bibfnamefont{G.}~\bibnamefont{Kirchmair}},
  \bibinfo{author}{\bibfnamefont{F.}~\bibnamefont{Zahringer}},
  \bibinfo{author}{\bibfnamefont{R.}~\bibnamefont{Gerritsma}},
  \bibinfo{author}{\bibfnamefont{M.}~\bibnamefont{Kleinmann}},
  \bibinfo{author}{\bibfnamefont{O.}~\bibnamefont{Guhne}},
  \bibinfo{author}{\bibfnamefont{A.}~\bibnamefont{Cabello}},
  \bibinfo{author}{\bibfnamefont{R.}~\bibnamefont{Blatt}}, \bibnamefont{and}
  \bibinfo{author}{\bibfnamefont{C.~F.} \bibnamefont{Roos}},
  \bibinfo{journal}{Nature} \textbf{\bibinfo{volume}{460}},
  \bibinfo{pages}{494} (\bibinfo{year}{2009}), \eprint{arXiv:0904.1655}.

\bibitem[{\citenamefont{Bartosik et~al.}(2009)\citenamefont{Bartosik, Klepp,
  Schmitzer, Sponar, Cabello, Rauch, and Hasegawa}}]{bartosik-2009-103}
\bibinfo{author}{\bibfnamefont{H.}~\bibnamefont{Bartosik}},
  \bibinfo{author}{\bibfnamefont{J.}~\bibnamefont{Klepp}},
  \bibinfo{author}{\bibfnamefont{C.}~\bibnamefont{Schmitzer}},
  \bibinfo{author}{\bibfnamefont{S.}~\bibnamefont{Sponar}},
  \bibinfo{author}{\bibfnamefont{A.}~\bibnamefont{Cabello}},
  \bibinfo{author}{\bibfnamefont{H.}~\bibnamefont{Rauch}}, \bibnamefont{and}
  \bibinfo{author}{\bibfnamefont{Y.}~\bibnamefont{Hasegawa}},
  \bibinfo{journal}{Phys. Rev. Lett.} \textbf{\bibinfo{volume}{103}},
  \bibinfo{pages}{040403} (\bibinfo{year}{2009}), \eprint{quant-ph/103.040403}.

\bibitem[{\citenamefont{Amselem et~al.}(2009)\citenamefont{Amselem, Radmark,
  Bourennane, and Cabello}}]{Amselem-independent}
\bibinfo{author}{\bibfnamefont{E.}~\bibnamefont{Amselem}},
  \bibinfo{author}{\bibfnamefont{M.}~\bibnamefont{Radmark}},
  \bibinfo{author}{\bibfnamefont{M.}~\bibnamefont{Bourennane}},
  \bibnamefont{and} \bibinfo{author}{\bibfnamefont{A.}~\bibnamefont{Cabello}},
  \bibinfo{journal}{Phys. Rev. Lett.} \textbf{\bibinfo{volume}{103}},
  \bibinfo{pages}{160405} (\bibinfo{year}{2009}), \eprint{quant-ph/0907.4494}.

\bibitem[{\citenamefont{Liu et~al.}(2009)\citenamefont{Liu, Huang, Gong, Sun,
  Zhang, Li, and Guo}}]{Liu-noncontextual}
\bibinfo{author}{\bibfnamefont{B.~H.} \bibnamefont{Liu}},
  \bibinfo{author}{\bibfnamefont{Y.~F.} \bibnamefont{Huang}},
  \bibinfo{author}{\bibfnamefont{Y.~X.} \bibnamefont{Gong}},
  \bibinfo{author}{\bibfnamefont{F.~W.} \bibnamefont{Sun}},
  \bibinfo{author}{\bibfnamefont{Y.~S.} \bibnamefont{Zhang}},
  \bibinfo{author}{\bibfnamefont{C.~F.} \bibnamefont{Li}}, \bibnamefont{and}
  \bibinfo{author}{\bibfnamefont{G.~C.} \bibnamefont{Guo}},
  \bibinfo{journal}{Phys. Rev. A} \textbf{\bibinfo{volume}{80}},
  \bibinfo{pages}{044101} (\bibinfo{year}{2009}).

\bibitem[{\citenamefont{Moussa et~al.}(2009)\citenamefont{Moussa, Ryan, Cory,
  and Laflamme}}]{moussa-2009}
\bibinfo{author}{\bibfnamefont{O.}~\bibnamefont{Moussa}},
  \bibinfo{author}{\bibfnamefont{C.~A.} \bibnamefont{Ryan}},
  \bibinfo{author}{\bibfnamefont{D.~G.} \bibnamefont{Cory}}, \bibnamefont{and}
  \bibinfo{author}{\bibfnamefont{R.}~\bibnamefont{Laflamme}},
  \emph{\bibinfo{title}{Testing contextuality on quantum ensembles with one
  clean qubit}} (\bibinfo{year}{2009}), \eprint{quant-ph/0912.0485}.

\bibitem[{\citenamefont{Mermin}(1990)}]{Mermin1990-KS}
\bibinfo{author}{\bibfnamefont{N.~D.} \bibnamefont{Mermin}},
  \bibinfo{journal}{Phys. Rev. Lett.} \textbf{\bibinfo{volume}{65}},
  \bibinfo{pages}{3373} (\bibinfo{year}{1990}).

\bibitem[{\citenamefont{Peres}(1990)}]{Peres1990-KS}
\bibinfo{author}{\bibfnamefont{A.}~\bibnamefont{Peres}},
  \bibinfo{journal}{Phys. Lett. A} \textbf{\bibinfo{volume}{151}},
  \bibinfo{pages}{107} (\bibinfo{year}{1990}).

\bibitem[{\citenamefont{Stairs}(1983)}]{Stairs-1983}
\bibinfo{author}{\bibfnamefont{A.}~\bibnamefont{Stairs}},
  \bibinfo{journal}{Philos. Sci} \textbf{\bibinfo{volume}{50}},
  \bibinfo{pages}{578} (\bibinfo{year}{1983}).

\bibitem[{\citenamefont{Heywood and Redhead}(1983)}]{HeywoodR1983}
\bibinfo{author}{\bibfnamefont{P.}~\bibnamefont{Heywood}} \bibnamefont{and}
  \bibinfo{author}{\bibfnamefont{M.~L.~G.} \bibnamefont{Redhead}},
  \bibinfo{journal}{Found. Phys.} \textbf{\bibinfo{volume}{13}},
  \bibinfo{pages}{481} (\bibinfo{year}{1983}).

\bibitem[{\citenamefont{{Cabello}}(2001)}]{Cabello2001}
\bibinfo{author}{\bibfnamefont{A.}~\bibnamefont{{Cabello}}},
  \bibinfo{journal}{Phys. Rev. Lett.} \textbf{\bibinfo{volume}{87}},
  \bibinfo{pages}{010403} (\bibinfo{year}{2001}),
  \eprint{arXiv:quant-ph/0101108}.

\bibitem[{\citenamefont{Pawlowski et~al.}(2009)\citenamefont{Pawlowski,
  Paterek, Kaszlikowski, Scarani, Winter, and Zukowski}}]{pawlowski-2009-461}
\bibinfo{author}{\bibfnamefont{M.}~\bibnamefont{Pawlowski}},
  \bibinfo{author}{\bibfnamefont{T.}~\bibnamefont{Paterek}},
  \bibinfo{author}{\bibfnamefont{D.}~\bibnamefont{Kaszlikowski}},
  \bibinfo{author}{\bibfnamefont{V.}~\bibnamefont{Scarani}},
  \bibinfo{author}{\bibfnamefont{A.}~\bibnamefont{Winter}}, \bibnamefont{and}
  \bibinfo{author}{\bibfnamefont{M.}~\bibnamefont{Zukowski}},
  \bibinfo{journal}{Nature} \textbf{\bibinfo{volume}{461}},
  \bibinfo{pages}{1101} (\bibinfo{year}{2009}), \eprint{arXiv.org:0905.2292}.

\bibitem[{\citenamefont{Lo et~al.}(2005)\citenamefont{Lo, Chau, and
  Ardehali}}]{LoChauArdehali2000}
\bibinfo{author}{\bibfnamefont{H.-K.} \bibnamefont{Lo}},
  \bibinfo{author}{\bibfnamefont{H.~F.} \bibnamefont{Chau}}, \bibnamefont{and}
  \bibinfo{author}{\bibfnamefont{M.}~\bibnamefont{Ardehali}},
  \bibinfo{journal}{J. Crypt.} \textbf{\bibinfo{volume}{18}},
  \bibinfo{pages}{133} (\bibinfo{year}{2005}), \eprint{quant-ph/0011056}.

\bibitem[{\citenamefont{Popescu and Rohrlich}(1997)}]{Popescu-Rohrlich}
\bibinfo{author}{\bibfnamefont{S.}~\bibnamefont{Popescu}} \bibnamefont{and}
  \bibinfo{author}{\bibfnamefont{D.}~\bibnamefont{Rohrlich}},
  \bibinfo{journal}{Phys. Rev. A} \textbf{\bibinfo{volume}{56}},
  \bibinfo{pages}{R3319} (\bibinfo{year}{1997}), \eprint{quant-ph/9610044}.

\bibitem[{\citenamefont{Conway and Kochen}(2007)}]{conway-2007}
\bibinfo{author}{\bibfnamefont{J.}~\bibnamefont{Conway}} \bibnamefont{and}
  \bibinfo{author}{\bibfnamefont{S.}~\bibnamefont{Kochen}},
  \emph{\bibinfo{title}{Thou shalt not clone one bit!}} (\bibinfo{year}{2007}),
  \eprint{quant-ph/0711.2310}.

\bibitem[{\citenamefont{Clauser et~al.}(1969)\citenamefont{Clauser, Horne,
  Shimony, and Holt}}]{CHSH}
\bibinfo{author}{\bibfnamefont{J.~F.} \bibnamefont{Clauser}},
  \bibinfo{author}{\bibfnamefont{M.~A.} \bibnamefont{Horne}},
  \bibinfo{author}{\bibfnamefont{A.}~\bibnamefont{Shimony}}, \bibnamefont{and}
  \bibinfo{author}{\bibfnamefont{R.~A.} \bibnamefont{Holt}},
  \bibinfo{journal}{Phys. Rev. Lett.} \textbf{\bibinfo{volume}{23}},
  \bibinfo{pages}{880} (\bibinfo{year}{1969}).

\bibitem[{\citenamefont{Tsirelson}(1980)}]{Tsirelson-bound}
\bibinfo{author}{\bibfnamefont{B.}~\bibnamefont{Tsirelson}},
  \bibinfo{journal}{Lett. Math. Phys.} \textbf{\bibinfo{volume}{4}},
  \bibinfo{pages}{93} (\bibinfo{year}{1980}).

\bibitem[{\citenamefont{{Gisin} et~al.}(2007)\citenamefont{{Gisin}, {Methot},
  and {Scarani}}}]{GisinMS2006-pseudo}
\bibinfo{author}{\bibfnamefont{N.}~\bibnamefont{{Gisin}}},
  \bibinfo{author}{\bibfnamefont{A.~A.} \bibnamefont{{Methot}}},
  \bibnamefont{and}
  \bibinfo{author}{\bibfnamefont{V.}~\bibnamefont{{Scarani}}},
  \bibinfo{journal}{Int. J. Quant. Inf.} \textbf{\bibinfo{volume}{5}},
  \bibinfo{pages}{525} (\bibinfo{year}{2007}), \eprint{arXiv:quant-ph/0610175}.

\bibitem[{\citenamefont{Csisz$\acute{a}$r and
  K{\"o}rner}(1978)}]{CsisarKorner_key_agreement}
\bibinfo{author}{\bibfnamefont{I.}~\bibnamefont{Csisz$\acute{a}$r}}
  \bibnamefont{and}
  \bibinfo{author}{\bibfnamefont{J.}~\bibnamefont{K{\"o}rner}},
  \bibinfo{journal}{IEEE} \textbf{\bibinfo{volume}{24}}, \bibinfo{pages}{339}
  (\bibinfo{year}{1978}).

\bibitem[{\citenamefont{Maurer}(1993)}]{Maurer_key_agreement}
\bibinfo{author}{\bibfnamefont{U.~M.} \bibnamefont{Maurer}},
  \bibinfo{journal}{IEEE} \textbf{\bibinfo{volume}{39}}, \bibinfo{pages}{773}
  (\bibinfo{year}{1993}).

\bibitem[{\citenamefont{Cabello}(2009)}]{cabello-2009}
\bibinfo{author}{\bibfnamefont{A.}~\bibnamefont{Cabello}},
  \emph{\bibinfo{title}{A bell inequality with local violation}}
  (\bibinfo{year}{2009}), \eprint{arxiv:0910.5507}.

\bibitem[{\citenamefont{Navascues et~al.}(2008)\citenamefont{Navascues,
  Pironio, and Acin}}]{navascues-2008-10}
\bibinfo{author}{\bibfnamefont{M.}~\bibnamefont{Navascues}},
  \bibinfo{author}{\bibfnamefont{S.}~\bibnamefont{Pironio}}, \bibnamefont{and}
  \bibinfo{author}{\bibfnamefont{A.}~\bibnamefont{Acin}}, \bibinfo{journal}{New
  J. Phys.} \textbf{\bibinfo{volume}{10}}, \bibinfo{pages}{073013}
  (\bibinfo{year}{2008}), \eprint{arXiv:0803.4290v1}.

\bibitem[{\citenamefont{DiVincenzo and Peres}(1997)}]{divincenzo-1997-55}
\bibinfo{author}{\bibfnamefont{D.~P.} \bibnamefont{DiVincenzo}}
  \bibnamefont{and} \bibinfo{author}{\bibfnamefont{A.}~\bibnamefont{Peres}},
  \bibinfo{journal}{Phys. Rev. A} \textbf{\bibinfo{volume}{55}},
  \bibinfo{pages}{4089} (\bibinfo{year}{1997}), \eprint{quant-ph/9611011}.

\bibitem[{\citenamefont{{Kretschmann} et~al.}(2008)\citenamefont{{Kretschmann},
  {Kribs}, and {Spekkens}}}]{kretschmann-2008-78}
\bibinfo{author}{\bibfnamefont{D.}~\bibnamefont{{Kretschmann}}},
  \bibinfo{author}{\bibfnamefont{D.~W.} \bibnamefont{{Kribs}}},
  \bibnamefont{and} \bibinfo{author}{\bibfnamefont{R.~W.}
  \bibnamefont{{Spekkens}}}, \bibinfo{journal}{Phys. Rev. A}
  \textbf{\bibinfo{volume}{78}}, \bibinfo{pages}{032330}
  (\bibinfo{year}{2008}), \eprint{arXiv:0711.3438}.

\bibitem[{\citenamefont{{Bechmann-Pasquinucci} and
  {Peres}}(2000)}]{BechmanPeres-2000}
\bibinfo{author}{\bibfnamefont{H.}~\bibnamefont{{Bechmann-Pasquinucci}}}
  \bibnamefont{and} \bibinfo{author}{\bibfnamefont{A.}~\bibnamefont{{Peres}}},
  \bibinfo{journal}{Phys. Rev. Lett.} \textbf{\bibinfo{volume}{85}},
  \bibinfo{pages}{3313} (\bibinfo{year}{2000}),
  \eprint{arXiv:quant-ph/0001083}.

\bibitem[{\citenamefont{Nagata}(2005)}]{nagata-2005-72}
\bibinfo{author}{\bibfnamefont{K.}~\bibnamefont{Nagata}},
  \bibinfo{journal}{Phys. Rev. A} \textbf{\bibinfo{volume}{72}},
  \bibinfo{pages}{012325} (\bibinfo{year}{2005}),
  \eprint{arXiv.org:quant-ph/0503158}.

\bibitem[{\citenamefont{{Svozil}}(2009)}]{Svozil}
\bibinfo{author}{\bibfnamefont{K.}~\bibnamefont{{Svozil}}}
  (\bibinfo{year}{2009}), \eprint{arXiv:0903.0231}.

\bibitem[{\citenamefont{{Brassard} et~al.}(2005)\citenamefont{{Brassard},
  {Broadbent}, and {Tapp}}}]{Brassard-pseudo}
\bibinfo{author}{\bibfnamefont{G.}~\bibnamefont{{Brassard}}},
  \bibinfo{author}{\bibfnamefont{A.}~\bibnamefont{{Broadbent}}},
  \bibnamefont{and} \bibinfo{author}{\bibfnamefont{A.}~\bibnamefont{{Tapp}}},
  \bibinfo{journal}{Found. Phys.} \textbf{\bibinfo{volume}{35}},
  \bibinfo{pages}{1877} (\bibinfo{year}{2005}),
  \eprint{arXiv:quant-ph/0407221}.

\bibitem[{\citenamefont{{Wehner}}(2006)}]{Wehner2005-bell}
\bibinfo{author}{\bibfnamefont{S.}~\bibnamefont{{Wehner}}},
  \bibinfo{journal}{Phys. Rev. A} \textbf{\bibinfo{volume}{73}},
  \bibinfo{pages}{022110} (\bibinfo{year}{2006}),
  \eprint{arXiv:quant-ph/0510076}.

\end{thebibliography}

\begin{appendix}
\section{Bound for probabilities of the first row: ideal PM box}
Here we shall prove the bound \eqref{eq:cons-q-gamma} in more detail.
Suppose  that upper row of Bob with certainty gives result $+++$.
This means, that upper observables from each Alice's row, 
have deterministic value $+1$. In other words, with certainty we have  $A_1''=+1,A_{2}''=+1,A_{3}''=+1$ (for notation, see Fig. \ref{fig:DPM-box}).
Thus, due to KS conditions, the two other observables in 
Alice's rows are correlated in two first rows, and anti-correlated in 
the last row:
\ben
&&A_{1}=A_{1}'\nonumber\\
&&A_{2}=A_{2}'\nonumber\\
&&A_{3}=-A_{3}'.
\label{eq:AA}
\een
On the other hand, from AB-correlations we have 
that all those observables are perfectly correlated with corresponding 
Bob's observables (i.e. $A_i=B_i$, $A'_i=B'_i$ and $A''_i=B''_i$ for all $i$).
Since each pair of observables in the  formula \eqref{eq:AA} is jointly measurable, we obtain  the following correlations in the total system:
\ben
&&A_{1}=B_{1},\quad A_{1}=B_{1}'\nonumber\\
&&A_{2}=B_{2},\quad A_{2}=B_{2}'\nonumber\\
&&A_{3}=B_{3},\quad A_{3}=-B_{3}'.
\label{eq:ABprim}
\een
There are three observables of Alice and six observables 
of Bob involved.  We now formulate Bell quantity
\ben
&&\gamma(A:B) \equiv \<A_1 B_1\> + \<A_2 B_2\>+\<A_3 B_3\>\nonumber \\
&&+\<A_1 B_1'\> +\<A_2 B_2'\>-\<A_3 B_3'\>.
\een
The correlations \eqref{eq:ABprim} mean that $\gamma(A:B)=6$. 
We shall now suppose that $\gamma$ is not necessarily 6, 
and derive constraints for Bob's probability distribution 
for the first row in terms of $\gamma$. 

To this end, we shall use another closely related Bell quantity:
\ben
&&\beta(A:B)\equiv p(A_1=B_1)+p(A_2=B_2)+p(A_3=B_3)\nonumber \\
&&+p(A_1=B_1')+p(A_2=B_2')+p(A_3\not=B_3').
\label{eq:beta-ap}
\een
Let us note that 
\be
\beta(A:B)=\frac12(\gamma(A:B)+6).
\ee
Note that from AB-correlations we have that 
\be
p(A_1=B_1)=p(A_2=B_2)=p(A_3=B_3)=1.
\ee
We shall now relate the other three probabilities with 
Bob's distribution of the first row. 
Let us denote Bob's distribution in the first row by $q_i,i=0,1,2,3$ with 
\ben
&&q_0=Pr(+1,+1,+1)\nonumber\\ 
&&q_1=Pr(+1,-1,-1)\nonumber\\
&&q_2=Pr(-1,+1,-1)\nonumber\\
&&q_3=Pr(-1,-1,+1).
\een
We have $q_0+q_1+q_2+q_3=1$. Consider now the marginal distributions 
of each observable from the row, and let $p_i$ be a probability 
that we obtain $+1$ for i-th observable in the row, 
i.e. $p_i=Pr(B_{i}''=+1)$. More generally we shall 
denote $p_i(k)=Pr(B_{i}''=k)$ for $k=\pm1$.
Here is relation between $q_i$ and $p_i$
\ben
&&p_1=q_0 + q_1\nonumber\\
&&p_2=q_0 + q_2\nonumber\\
&&p_3=q_0 + q_3.
\een
Note that $p_i$'s do not sum up to 1, as they represent three 
separate probability distributions $(p_i,1-p_i)$.  The above relation gives 
\ben
&&q_0= \frac12 (-1+ p_1 +p_2+p_3)\nonumber \\
&&q_1= \frac12(1+ p_1 -p_2-p_3)\nonumber \\
&&q_2= \frac12(1- p_1 +p_2-p_3)\nonumber\\
&&q_3= \frac12(1- p_1 -p_2+p_3).
\label{eq:qvsp}
\een
At first sight, it might seem that the above relation 
allow for negative $q_i$'s, but we have to recall, 
that not all $p_i$'s are allowed due to KS conditions.
We shall now first derive the constrains for $p_i$'s 
and then  translate them into the constraints for $q_i$'s.

Now, note first that due to AB-correlations, each Alice's observable in the first row
have the same distribution  as Bob's corresponding observable in his first row. 
Thus $p_i$'s are equal also to marginal distributions 
of observables from Alice's first row. Consider now a chosen Alice's 
observable from the first row. Due to KS conditions we have 
\be
Pr(A_1=A_1')=Pr(A_{1}''=+1)= p_1.
\ee
Similarly, we have 
\ben
&&Pr(A_2=A_2')=p_2\nonumber\\
&&Pr(A_3=A_3')=p_3.
\een
We consider now three events: $X=\{A_1=A_1'\}$, $Y=\{A_1'=B_1'\}$ and $Z=\{A_1=B_1'\}$.
Clearly $X\cap Y\subset Z$, and using an elementary inequality valid for any events 
\be 
Pr(X\cap Y)\geq Pr(X)+Pr(Y) -1
\label{eq:prob-el}
\ee
we obtain that $Pr(Z)\geq Pr(X)+Pr(Y) -1$ 
which means that 
\be
Pr(A_1=B_1')\geq p_1, 
\ee
since $Pr(Y)=1$ (from perfect AB-correlations).
Similarly we obtain 
\ben
&&Pr(A_2=B_2')\geq p_2\nonumber\\
&&Pr(A_3\not=B_3')\geq p_3.
\een
Thus we obtain the following relation between $\beta$ and 
$p_i$'s for perfect correlations:
\be
\beta\geq p_1+p_2+p_3 +3.
\label{eq:cons-beta}
\ee

Thus we have proved the following lemma: 
\begin{lemma}
Let $\beta$ denote the Bell quantity \eqref{eq:beta-ap}, and let $p_i$ 
be probabilities of obtaining $+1$ while measuring $i$-th observable 
of the first row of Bob. Then 
\be
p_1+p_2+p_3 \leq \beta-3.
\ee
\end{lemma}

The lemma was obtained by analysing a situation when Bob receives 
output $+++$ in the first row with some probability, 
not necessarily equal to 1.  In a similar way 
we can treat three other events: $+--$, $-+-$ and $--+$. 
They lead again to prefect correlations/anticorrelations 
for the observables \eqref{eq:ABprim}. 
For $+--$,  $A_2$ and $B_2'$, are anticorrelated, 
for $-+-$,  $A_1$ and $B_1'$ are anticorrelated and finally 
for $--+$  $A_i$ are anticorrelated with $B_i'$ for $i=1,2,3$. 
This leads, in turn, to three other Bell quantities $\beta_1$, $\beta_2$ and $\beta_3$
which are all equivalent to the quantity $\beta$,
via redefining some observables (multiplying by $-1$). 
Thus maximal quantum mechanical values of all those quantities are equal.
Let us call this maximal value $\beta_0$

Analysing the mentioned three cases in the same way as above, we obtain 
\ben
&& p_1+ (1-p_2)+(1-p_3) \leq \beta_0-3\nonumber\\
&& (1-p_1)+ p_2+(1-p_3) \leq \beta_0-3\nonumber\\
&& (1-p_1)+ (1-p_2)+p_3 \leq \beta_0-3.
\een
Using \eqref{eq:qvsp},  these inequalities and the fourth one 
obtained in the lemma above, we get 
\be
q_i\leq \frac12(\beta_0-4),
\label{eq:bound-q-beta}
\ee
and translating it into $\gamma_0$  (which is a maximal 
quantum value of quantity $\gamma$) we obtain 
the following lemma:
\begin{lemma}
The prefect AB-correlations and perfect KS conditions 
imply that 
\be
q_i\leq \frac14(\gamma_0-2),
\ee
where $\{q_i\}$ is the joint probability distribution of 
the outcomes of Bob's first row. 
\end{lemma}

\section{Bound for probabilities of the first row: noisy PM box}

In this section we shall prove inequalities \eqref{eq:hbe} and \eqref{eq:hba},
as well as provide the estimate for the noise threshold.
We consider a noisy PM box. The KS conditions are still perfect,
because in every row (column) one of observables is not 
measured, but is produced to fit the conditions. Thus only 
the AB-correlations are not perfect. 

The error estimation in the protocol can be divided into 
two parts. First, Alice measures
chosen at random columns and Bob - chosen at random rows. 
There are 9 different combinations of rows and columns,
and in each combination there is one common observable. 
We shall assume that for each combination, there is the same probability 
of error $\epsilon$ (i.e. of obtaining different outcomes). 
This stage determines bounds for probabilities of Bob's row
(hence also puts bounds on $H(B|E)$).  

In the second part of the error estimation, Alice measures first row, and Bob measures first row too. Let the probability that 
their outcomes disagree be $\tep$. 
In order to have a single noise parameter, 
we need to relate $\tep$ with error probability for single nodes of the row, 
which for simplicity we also assume to be $\epsilon$. From KS condition 
we get 
\be
\ep=\frac23 \tep.
\label{eq:epep}
\ee
This part of error estimation will put bound on $H(B|A)$, 
which we shall do in next section, where we derive 
bounds for both conditional entropies.  

In this section we shall deal with the  first part.
Let us name the observables of the first Alice's row $A''_i$, 
and first Bob's row $B''_i$. 

We want to find constraints for probabilities of Bob's first row
$P(B'')$. We shall start with $P(B''_i=+1)\equiv p_i$, 
By inequality \eqref{eq:prob-el}, and using the fact that 
$B''_i$ and $A''_i$ are correlated with probability $1-\ep$ we have 
\be
P(A''_i=+1)\geq p_i+(1-\ep) -1=p_i-\ep
\ee
By KS condition 
\ben
&&P(A_1=A'_1)=P(A''_1=+1)\nonumber \\
&&P(A_2=A'_2)=P(A''_2=+1)\nonumber\\
&&P(A_1\not=A'_1)=P(A''_3=+1).
\een
Again, since $B'_i$ and $A'_i$ are correlated with probability $1-\ep$
we have 
\ben
&P(A_1=B_1')\geq P(A_1=A'_1)+P(A'_1=B'_1)-1\geq p_1 - 2\ep&\nonumber \\
&P(A_2=B_2')\geq P(A_2=A'_2)+P(A'_2=B'_2)-1\geq p_2 - 2\ep&\nonumber \\
&P(A_3\not=B_3')\geq P(A_3=A'_3)+P(A'_3=B'_3)-1\geq p_3 - 2\ep&\nonumber\\
\een
On the other had we have 
\be
P(A_i=B_i)\geq 1-\ep
\ee
so that 
\be
\beta \geq p_1+p_2+p_3 +3 - 9\ep
\ee
i.e. 
\be
p_1+p_2+p_3 \leq \beta'-3
\ee
i.e. we have obtained constraints of the same form as in the 
noiseless case \eqref{eq:cons-beta}, with $\beta $ replaced with 
$\beta'=\beta + 9\ep$. Similarly we can proceed for 
three other cases in Bob's first row $(+--,-+-,--+)$
to obtain the following bound 
for joint probabilities $q_i$ of outcomes of the first Bob's row:
\be
q_i\leq \frac12(\beta_0 +9 \ep -4),
\label{eq:bound-q-betaprim}
\ee
and finally:
\be
q_i\leq \frac14(\gamma_0-2)+\frac{9\ep}{2}.
\ee
Due to our bound $\gamma_0\leq 5.6364$ we get 
\be
q_i\leq 0.9091+ 4.5 \ep 
\ee
For all $i=0,1,2,3$.

\section{Bound on conditional entropies}
Let us start with bound for $H(B|A)$.
This entropy is evaluated on joint probability distribution
coming from first row on Alice side and first row on Bob's side. 
Let $\tep$ be probability of error (i.e. that the outcomes differ).
Then we use Fano's inequality:
\be
H(B|A)\leq h(\tep) + \tep \log(|B|-1)=h(\tep) + \tep \log 3
\ee
A more natural error parameter for the whole protocol is probability of error 
in single node. We assume that in each node the probability of error is the same, 
equal to $\ep$, hence by \eqref{eq:epep} 

\be
H(B|A)\leq h\biggl(\frac32 \ep\biggr)+\frac32 \ep \log 3.
\ee

Let us now estimate conditional entropy $H(B|E)$. 
First, consider a box which satisfies KS conditions and 
have probability of anticorrelations $\ep$. 
The entropy of Bob's first row is bounded from below by 
\be
H(B;\ep)\geq h(x)\equiv f(\ep)
\ee
where $x=\min(0.9091 + 4.5\ep,1)$. 
Note that $f$ is nonnegative, decreasing function of $\ep$. 
Let Eve make
a measurement on her system. This splits Alice's and Bob's box into ensemble:
$R_{AB}=\sum_e r_e R^e_{AB}$. For notational convenience, let us use indices $i$ in place of $e$. Then 
\be
H(B|E)=\inf \sum_i r_i H(B)_i
\ee
where $H(B)_i$ is Bob's first row entropy of box $R^i_{AB}$ and
infimum is taken over all decompositions $R_{AB}=\sum_i r_i R^i_{AB}$
The new boxes $R^e$ satisfy 
\be
\sum_i r_i \ep_i=\ep.
\ee
Thus $H(B|E)$ is bounded as follows
\be
H(B|E)=\geq \inf_{\{r_i,\ep_i\}} \sum_i r_i f(\ep_i),
\label{eq:hbe-ap}
\ee
where  $\sum_i r_i\ep_i=\ep$, $\sum_i r_i =1$. 
To estimate the above quantity note that by Markov inequality 
we have 
\be
\sum_{i:\ep_i<\delta} r_i \geq 1-\frac{\ep}{\delta}.
\ee
This gives 
\be
H(B|E)\geq \sup_\delta(1-\frac{\ep}{\delta})h(0.9091 +4.5\delta).
\ee

Overall, we obtain the following bound on key rate:
\ben
&&K\geq H(B|E)- H(B|A) \geq \sup_\delta(1-\frac{\ep}{\delta})
h(0.9091 +4.5\delta) - \nonumber\\
&&\biggl\{h\biggl(\frac32 \ep\biggr)+\frac32 \ep \log 3\biggr\}. 
\een
Putting $\delta=1.8 \ep$ we obtain the following noise threshold, below 
which sharing key is possible. 
\be
\ep_0 \leq 0.68\%.
\ee
This is smaller than typical thresholds obtained 
from CHSH, which are of order of $2\%$. However we have not performed 
optimization in \eqref{eq:hbe-ap}, which would give a better rate. 

\section{Bound for a Bell inequality}
Here we outline the way we obtained quantum-mechanical bound for the Bell inequality
\eqref{eq:Bell-gamma}.
We shall follow Refs. \cite{Wehner2005-bell} and \cite{navascues-2008-10}
(their methods originate from  Tsirelson approach \cite{Tsirelson-bound}).
Namely, a matrix $\tilde \Gamma$ with the following matrix elements $\tilde\Gamma_{ij} =\<\psi|X_i^\dagger X_j|\psi\>$ 
is always positive semidefinite, for any state $\psi$ and  any collecion of 
operators $X_i$. Hence a matrix $\Gamma=\frac12(\tilde\Gamma+\tilde\Gamma^*)$ 
where $*$ denotes complex conjugation is a real positive semidefinite matrix. 

For our purpose the role of $X_j$'s will be played by the following ten operators
\be
I,A_1,A_2,A_3,B_1,B_2,B_3,B_1',B_2',B_3'.
\ee 
The resulting matrix $\Gamma$ has thus $1$'s on diagonal, 
and moreover satisfies a couple of constraints. Namely, 
The equalities 
\be
\<I B_1\>=\<B_2 B_3\>,\quad\<I B_2\>=\<B_1 B_3\>,\quad \<I B_3\>=\<B_1 B_2\>
\ee
imply 
\be
\Gamma_{1,5}=\Gamma_{6,7},\quad\Gamma_{1,6}=\Gamma_{5,7},
\quad\Gamma_{1,7}=\Gamma_{6,5}.
\label{eq:con1}
\ee
Similar equalities for $B'_i$'s imply further three constraints
\be
\Gamma_{1,8}=\Gamma_{9,10},\quad\Gamma_{1,9}=\Gamma_{8,10},
\quad\Gamma_{1,10}=\Gamma_{8,9}.
\label{eq:con2}
\ee 
Here is the full matrix $\tilde \Gamma$, where by $\times$ 
we denote undetermined elements, which are 
constrained only by its positivity.  
The elements $b_i$ and $b_i'$ are also undetermined. we do not 
write explicitly elements below the diagonal, 
as the matrix is Hermitian.
\begin{widetext}
\be
\bea{c|cccccccccc}
       &I&\<A_1\>&\<A_2\>&\<A_3\>&\<B_1\>&\<B_2\>&\<B_3\>&\<B_1'\>&\<B_2'\>&\<B_3'\>\\
       \hline
I       &1 &\t&\t&\t&b_1        &b_2      &b_3       &b_1'        &b_2'        &b_3'\\
\<A_1\> &  &1 &\t&\t&\<A_1 B_1\>&\t       &\t        &\<A_1 B_1'\>&\t          &\t  \\
\<A_2\> &  &  & 1&\t&\t        &\<A_2B_2\>&\t        &\t          &\<A_2B_2'\> &\t  \\
\<A_3\> &  &  &  & 1&\t        &\t        &\<A_3B_3\>&\t          & \t  &\<A_3B_3'\>\\
\<B_1\> &  &  &  &  &1         &b_3       &b_2       &\t          &\t          &\t  \\
\<B_2\> &  &  &  &  &          &1         &b_1       &\t          &\t          &\t  \\
\<B_3\> &  &  &  &  &          &          &1         &\t          &\t          &\t \\
\<B_1'\>&  &  &  &  &          &          &          &1           &b_3'       &b_2' \\
\<B_2'\>&  &  &  &  &          &          &          &            &1           &b_1'\\
\<B_3'\>&  &  &  &  &          &          &          &            &            &1  \\
\eea
\ee
\end{widetext}

In terms of the matrix $\Gamma$, the Bell's inequality reads as follows:
\be
\gamma=\frac12 \tr(\Gamma W),
\label{eq:sdp}
\ee
where 
\be
W=\left[\bea{cccccccccc}
0&0&0&0&0&0&0&0&0&0\\
0&0&0&0&1&0&0&1&0&0\\
0&0&0&0&0&1&0&0&1&0\\
0&0&0&0&0&0&1&0&0&-1\\
0&1&0&0&0&0&0&0&0&0\\
0&0&1&0&0&0&0&0&0&0\\
0&0&0&1&0&0&0&0&0&0\\
0&1&0&0&0&0&0&0&0&0\\
0&0&1&0&0&0&0&0&0&0\\
0&0&0&-1&0&0&0&0&0&0\\
\eea\right]
\ee
Now, we obtain upper bound for value $\gamma$, 
by maximizing the right-hand-side of \eqref{eq:sdp}
under the constraints \eqref{eq:con1} and \eqref{eq:con2}  
and $\Gamma\geq 0$. This can be done by a standard SDP packages. 
We have used SDPT3 package for Matlab. However, it turns out 
that the upper bound here is trivial, i.e. it is equal to $6$. 
This can be directly verified: namely, there exists a matrix 
which gives 6, is positive and satisfies the constratins:
\be
\Gamma_0=\left[\bea{cccccccccc}
1&0&0&0&0&0&0&0&0&0\\
0&1&0&0&1&0&0&1&0&0\\
0&0&1&0&0&1&0&0&1&0\\
0&0&0&1&0&0&1&0&0&-1\\
0&1&0&0&1&0&0&1&0&0\\
0&0&1&0&0&1&0&0&1&0\\
0&0&0&1&0&0&1&0&0&-1\\
0&1&0&0&1&0&0&1&0&0\\
0&0&1&0&0&1&0&0&1&0\\
0&0&0&-1&0&0&-1&0&0&1\\
\eea\right]
\ee
Thus the application of the so called first SDP hierarchy 
according to terminology of \cite{navascues-2008-10} does not result in 
a nontrivial upper bound. We have therefore checked the 
second hierarchy, where $X_i$'s are all possible products of pairs 
of the set $I,A_1,A_2,A_3,B_1,B_2,B_3,B_1',B_2',B_3'$. 
This leads to another SDP program, which produces a nontrivial 
upper bound for the Bell's inequality, i.e. 5.6364.
The set of constraints of type \eqref{eq:con1} we have generated on Mathematica,
while the SDP program was run by use of SDPT3 free tool for Matlab.

\end{appendix}

\end{document}